\documentclass[pra,amsmath,amsfonts,amssymb,twocolumn,showpacs]{revtex4}

\usepackage[utf8]{inputenc}
\usepackage[T1]{fontenc}

\usepackage{graphicx}
\usepackage{epsfig}
\usepackage{psfrag}

\usepackage{amsmath}
\usepackage{amssymb}
\usepackage{amsfonts}
\usepackage{latexsym}
\usepackage{theorem}
\usepackage{bm}
\usepackage{psfrag}

\newtheorem{theo}{Theorem}

{\hfill$\blacksquare$\par\vskip12ptplus60pt}

\def\ic[#1]{{\tt [#1]}} 

\newcommand{\ket}[1]{\ensuremath{|#1 \rangle}}
\newcommand{\bra}[1]{\ensuremath{\langle #1|}}
\newcommand{\braket}[2]{\ensuremath{\langle #1 | #2 \rangle}}

\newcommand{\ketbra}[2]{\ensuremath{| #1 \rangle \! \langle #2 |}}
\newcommand{\kettbra}[1]{\ketbra{#1}{#1}}

\def\idty{{\leavevmode\rm 1\mkern -5.4mu I}}
\def\Cx{\ensuremath{\mathbb{C}}}
\def\Rl{\ensuremath{\mathbb{R}}}

\newcommand{\tr}{\ensuremath{{\rm tr}}}

\newcommand{\hh}{{\mathcal{H}}}
\newcommand{\kk}{{\mathcal{K}}}
\newcommand{\rr}{{\mathcal{R}}}
\def\rrh{\hbox{$\widehat{\mskip4mu\rr}\mskip2mu$}}

\def\bs#1#2{\Phi_{#1}(#2)}
\def\bsh#1#2{\widehat\Phi_{#1}(#2)}

\begin{document}

\title{A Meaner King uses Biased Bases}

\author{Michael Reimpell}
\author{Reinhard F.~Werner}

\affiliation{Institut f\"ur Mathematische Physik,
Technische Universit\"at Braunschweig,
Mendelssohnstra{\ss}e 3, D-38106 Braunschweig, Germany}

\date{\today}

\begin{abstract}
The mean king problem is a quantum mechanical retrodiction problem,
in which Alice has to name the outcome of an ideal measurement on a
$d$-dimensional quantum system, made in one of $(d+1)$ orthonormal
bases, unknown to Alice at the time of the measurement. Alice has to
make this retrodiction on the basis of the classical outcomes of a
suitable control measurement including an entangled copy. We show
that the existence of a strategy for Alice is equivalent to the
existence of an overall joint probability distribution for $(d+1)$
random variables, whose marginal pair distributions are fixed as the
transition probability matrices of the given bases. In particular,
for $d=2$ the problem is decided by John Bell's classic inequality
for three dichotomic variables. For mutually unbiased bases in any
dimension Alice has a strategy, but for randomly chosen bases the
probability for that goes rapidly to zero with increasing $d$.
\end{abstract}

 \maketitle

\section{Introduction}
The mean king problem was first introduced by Vaidman et al.\
\cite{Vaid}, and has since received a lot of attention as a basic
quantum mechanical retrodiction problem: In the story, physicist
Alice faces the mean king, who asks her to prepare a quantum system,
on which his men will perform a von Neumann measurement in one of a
specified set of orthonormal bases. Alice is not present during this
measurement, and knows neither the basis chosen nor the result
obtained. She is then allowed a final check on the system (typically
including entangled records of the initial preparation), leaving her
with some classical measurement values only. She is then told which
basis was used and is asked to correctly name the values found by
the king's men.

In \cite{Vaid} the system was a qubit, and the three bases involved
were the eigenbases of the three Pauli matrices. These are a special
case of {\it mutually unbiased bases} (``MUBs''), which means that
after preparing a basis state of any of these bases, the probability
distributions in all other bases will be uniform. It was
subsequently shown that a maximal set of $(d+1)$ mutually unbiased
bases exist for a $d$-dimensional Hilbert space, whenever $d$ is a
power of a prime, but to decide the existence of such a set for
other dimensions (e.g., $d=6$) has proved to be very hard. A status
report on this problem, which has also many repercussions on other
problems in quantum information is to be found in \cite{problem13}.

Note, however, that the basic statement of the problem makes no
reference to the MUB property of the bases chosen by the king. So,
supposing the game is to be played in $d=6$, why should the king
make things difficult (if not impossible) for himself by trying to
find first a set of mutually unbiased bases in that dimension? Why
not just pick any bases at random, say? Moreover, as we will see
below, in the mutually unbiased case, Alice has a very simple way to
compute a safe strategy. Again, a really mean king might make things
more difficult for Alice here. These remarks go against another
intuition, which would seem to make the choice of unbiased bases the
meanest option for the king: Indeed, if Alice were trying to just
make a measurement in one of the king's bases, hoping to pick the
right one, in all but the lucky case her results will be totally
useless. If the game were to be played many times, and Alice's aim
was to be right as often as possible, she could improve her guesses
using the correlations between bases. This gain is nullified in the
MUB case. But the problem is not set like that: We demand of a
solution that Alice is right {\it in every single run}, and this is
not made easier in the least by the existence of some statistical
correlations between the bases. In other words, the intuition that
unbiased bases are an especially mean choice by the king is
fallacious.

We therefore drop the assumption of unbiasedness and ask, for {\it
any} choice of finitely many bases by the king: Can Alice find a
strategy, consisting of an initial entangled preparation and a
suitable measurement on the joint system after the kings men are
through with their part, such that she gets the right value with
probability one?

Not very much has been done about this problem without assuming
mutual unbiasednes. Some special cases have been discussed in
\cite{haya1,haya2,benmen,Ozawa}. However, the available studies
apparently remain incomplete even in the qubit case ($d=2$).

\section{Summary of results}
\label{sec:results}

To state our main results, let us fix some notation for the rest of
the paper. The system Hilbert space on which the king's men make
their measurement will be denoted by $\hh$, and has dimension $d$.
The number of bases chosen will be $k$, and the bases themselves
will be  denoted by $\bs bi$, for $b=1,\ldots,k$ and $i=1,\ldots,d$.
An important property of a choice of bases is the space $\rr$ of
hermitian operators spanned by all the $\kettbra{\bs bi}$. This
space describes how many density operators we can distinguish with
measurements in the given bases. Since $\sum_i\kettbra{\bs
bi}=\idty$ for any basis, we expect only $(d-1)$-dimensions giving
new information for each basis, so together with the identity we
expect $\rr$ to be $(k(d-1)+1)$-dimensional. If this number is
achieved, we will call the chosen basis set {\it non-degenerate}. Of
course, $\dim\rr$ cannot exceed the dimension $d^2$ of the space of
all hermitian operators on $\hh$, so for $k>(d+1)$ every choice of
bases is degenerate in this sense. The interesting property in this
case is that $\dim\rr=d^2$, i.e., that the set of bases is {\it
tomographically complete}. Of course, for $k=(d+1)$, which is the
standard case, non-degeneracy and tomographic completeness are the
same property.

For any pair of bases, the values
\begin{equation}\label{probs}
    p_{bc}(i,j)= \frac1d\,\left|\braket{\bs bi}{\bs cj}\right|^2
\end{equation}
are the joint probabilities of a pair of $d$-valued random
variables, each of which is uniformly distributed. We say that a
collection of $k$ bases {\it admits a classical model}, if these
probabilities are marginals of some joint distribution of all $k$
variables (each taking $d$ values). Since this property only
involves the absolute values of scalar products, not their phases,
and therefore captures only a small part of the information about
the relative position of the bases, it is perhaps rather unexpected
that the existence of a classical model is very closely linked to
the existence of Alice's strategy. This is described in the
following Theorem, our main result:

\begin{theo}Let $\{\bs bi\}$ be a collection of $k$ orthonormal
bases in a $d$ dimensional Hilbert space. Then\\
(1) if the bases are non-degenerate (in particular, $k\leq(d+1)$)
and the bases admit a classical model, then Alice can find a safe
strategy in the mean king's problem with these
bases.\\
(2) Conversely, if the set of bases is tomographically complete (in
particular, $k\geq(d+1)$), and
if Alice has a strategy, then the bases allow a classical model.\\
(3) In the case (1), Alice's strategy may begin with a maximally
entangled state, and in the case (2) a pure initial state is
necessarily maximally entangled.
\end{theo}

Before going into the proof, let us see what this Theorem says about
some basic examples.

\subsection{Mutually unbiased bases} By definition a set of $k$
bases in $d$ dimensions is mutually unbiased if, with the notation
from (\ref{probs}), we have
\begin{equation}\label{mubcond}
 p_{bc}(i,j)= \delta_{bc}\delta_{ij}\frac1d+(1-\delta_{bc})\frac1{d^2}.
\end{equation}
From this a classical model is obvious, namely $k$ {\it
statistically independent} uniformly distributed random variables.
In order to compute the dimension of the span of the $\kettbra{\bs
bi}$, let us take the $(kd)\times(kd)$-matrix of Hilbert Schmidt
scalar products (defined for operators $A,B$ by $\langle
A|B\rangle_{HS}:=\tr (A^*B)$) of these vectors, which is just the
expression (\ref{mubcond}), interpreted as a matrix $M_{bi,cj}$. Its
rank is the dimension we are looking for, and easily computed as
$k(d-1)+1$, by determining all eigenvalues of $M$. Hence MUBs are
non-degenerate for all $k\leq(d+1)$, and for any number of MUBs the
mean king can come up with, Alice has a strategy. This result was
previously obtained by another method in \cite{Ozawa}.

\subsection{Qubits}
\begin{figure}[t]
\psfrag{p}[c][c][1][45]{$p_{ab}(1,1)$}
\psfrag{q}[c][c][1][-30]{$p_{ca}(1,1)$}
\psfrag{r}[c][c][1][-85]{$p_{bc}(1,1)$}
\includegraphics*[width=0.8\columnwidth]{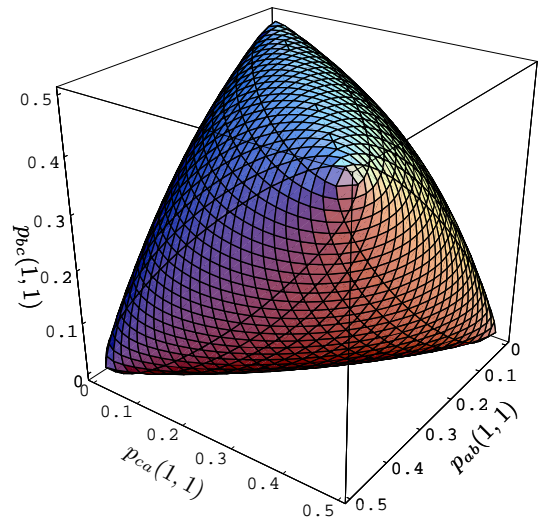}
\caption{Possible range of triples
   $\bigr(p_{ab}(1,1),p_{bc}(1,1),p_{ca}(1,1)\bigl)$.
   The range of triples admitting a classical model is
   described as tetrahedron inside this body, with the same corners.}
\label{fig:qubitcase}
\end{figure}
Another interesting special case, discussed in \cite{haya1}, is
$d=2$, $k=3$. Choosing a basis in $d=2$ is the same as choosing a
pair of antipodal points on the Bloch sphere. Three bases are
tomographically complete iff these points do not lie in a plane. The
existence of a classical model in this case is one of the ancestral
problems of quantum information theory, namely precisely the
existence of such models for three dichotomic variables
characterized by Bell's original three-variable inequality
\cite{Bell}. The joint distribution (\ref{probs}) belonging to two
bases $b,c$ is characterized (for $d=2$) by the single number
$p_{bc}(1,1)$. Fig.~\ref{fig:qubitcase} shows the possible range of
triples $\bigr(p_{ab}(1,1),p_{bc}(1,1),p_{ca}(1,1)\bigl)$. The range
of triples admitting a classical model, and hence a safe strategy
for Alice, is described by Bell's inequalities as the tetrahedron
inside this body. If the bases are chosen independently and with
unitarily invariant distribution (Haar measure), the probability for
this subset is exactly $1/3$. This can be computed analytically by
reducing it to a problem of three independent uniformly distributed
vectors on the Bloch sphere.

\section{Proof of Main Result}
In the first round, Alice chooses a Hilbert space $\kk$ and prepares
a density operator $\rho$ on $\hh\otimes\kk$. The first system is
left to the king's men, who perform their von Neumann measurement in
one of the bases $\Phi_b$, leaving a state $(\kettbra{\bs
bi}\otimes\idty)\rho(\kettbra{\bs bi}\otimes\idty)$, conditional on
their measured result being $i$. Finally, Alice will make a
measurement on $\hh\otimes\kk$, with some outcomes $x\in X$. This is
described by positive operators $F_x$ on $\hh\otimes\kk$, with
$\sum_xF_x=\idty$. The precise nature of the outcomes is irrelevant.
All that counts is that the value $x$ provides Alice with a rule
what to answer, if the king discloses that basis $b$ was used by his
men. We can express this by introducing a ``guessing function'', but
we might just as well take the rule itself as the outcome (possibly
grouping together some outcomes leading to the same guesses). Hence
we choose the outcome set
\begin{equation*}
X = \{1,\ldots,d\}^k\ =
\bigl\{x:\{1,\ldots,k\}\to\{1,\ldots,d\}\bigr\}
\end{equation*}
with the interpretation that $x(b)$ is the answer Alice will give,
if her measurement gave the value ``\,$x$\,'', and the King
discloses $b$. The requirement that she is right every time is the
basic equation for $\rho$ and $F_x$ we have to solve:
\begin{equation} \label{basiceq}
\begin{split}
 & \tr\Bigl(\rho (\kettbra{\bs bi}\otimes\idty)
          F_x(\kettbra{\bs bi}\otimes\idty)
    \Bigr)\\
& = \lambda_{i,x} \delta_{i,x(b)}, \quad \lambda_{i,x} \geq 0, \quad
\sum_{i,x} \lambda_{i,x} = 1.
\end{split}
\end{equation}

At this point we can make the first simplifications. Suppose, for
example, that Alice has found a solution using a mixed state $\rho$,
and that $\Psi$ is some unit vector in the support of $\rho$, so
that $\kettbra\Psi\leq\lambda\rho$ for some positive $\lambda$. Then
after replacing $\rho$ by $\kettbra\Psi$, the zeros in
(\ref{basiceq}) will still all be in the right places, and since we
chose $\Psi$ as a unit vector, we have found a solution with a pure
initial state $\rho=\kettbra\Psi$. Next, we write $\Psi$ as
\begin{equation} \label{Som}
\Psi = (\idty\otimes S)\Omega \quad \text{with } \Omega =
\sum_{\alpha=1}^d \ket{\alpha\alpha}\in\hh\otimes\hh.
\end{equation}
$\Omega$ is the maximally entangled vector, and $S$ is an operator,
whose matrix elements in a suitable basis of $\kk$ are the vector
components of $\Psi$, normalized so that $\tr(S^*S)=1$. Then in
(\ref{basiceq}) we can commute $S$ past the projections acting only
on the first factor, and simplify the expression by introducing the
vectors
\begin{eqnarray}\label{bshat}
    \bsh bi&=&(\kettbra{\bs bi}\otimes\idty)\Omega
             \nonumber\\
        &=&\bs bi\otimes\overline{\bs bi}
    \quad\in\hh\otimes\hh,
\end{eqnarray}
where the bar indicates componentwise complex conjugation in the
basis $\ket\alpha$, in which $\Omega$ takes the form (\ref{Som}).
Then the basic equation (\ref{basiceq}) becomes
\begin{equation}\label{basiceq1}
    \bra{\bsh bi}(\idty\otimes S)^*F_x(\idty\otimes S)\ket{\bsh bi}
    = \lambda_{i,x} \delta_{i,x(b)}\;.
\end{equation}
Now suppose $\eta$ is a vector in the support of the operator in
this bracket. Then substituting $\kettbra\eta$ for the operator will
still give zero, whenever $i\neq x(b)$, and hence
\begin{equation}\label{eta1}
    \braket{\bsh bi}\eta=0
    \quad\mbox{if}\quad i\neq x(b).
\end{equation}
But also the scalar products for $i=x(b)$ are essentially fixed: We
have $\sum_i\kettbra{\bs bi}=\idty$ for any basis $b$, which
translates to $\sum_i\bsh bi=\Omega$ via (\ref{bshat}). Therefore,
we can sum (\ref{eta1}) over $i$, obtaining
\begin{equation}\label{eta2}
    \braket{\bsh bi}\eta=\braket\Omega\eta\ \delta_{i,x(b)}.
\end{equation}
We will call such vectors {\it safe vectors} for Alice (and the
particular outcome $x$).

\subsection{Structure of safe vectors}
How many safe vectors can Alice find? A key role for answering this
question is played by the space $\rr$ introduced in
Sect.~\ref{sec:results}, or, equivalently by its image
$\rrh\subset\hh\otimes\hh$ under the identification of operators on
$\hh$ and elements of $\hh\otimes\hh$:
\begin{equation}\label{rrh}
   \rrh={\rm lin}_\Rl \{\bsh bi\},
\end{equation}
its complex linear span $\rrh_\Cx$, and its orthogonal complement
$\rrh^\perp$. For every $x\in X$ we arrive at the following 
alternative: It may happen that there is no vector $\eta$
satisfying (\ref{eta2}) with $\braket\Omega\eta\neq0$. Then all
solutions of that equation are in $\rrh^\perp$, which also means
that such values $x$ can never occur as a result of Alice's
measurement. Alice's strategy will have to rely on the other cases,
i.e., the subset of those $x\in X$, for which a non-trivial solution
$\eta$ of (\ref{eta2}) exists. To get a standard solution, we
multiply $\eta$ with a scalar so that $\braket\Omega\eta=1$.
Moreover, we can apply to $\eta$ the orthogonal projection to
$\rrh_\Cx$, thus obtaining a solution which is {\it uniquely}
determined, since all scalar products with vectors from this space
are fixed. We note that since all its scalar products with the $\bsh
bi$ are real, we can even conclude that $\eta_x\in\rrh$. Hence
whenever a non-zero solution exists for some $x\in X$, we can pick a
unique solution $\eta_x$, determined by the conditions
\begin{equation}\label{etax}
    \braket{\bsh bi}{\eta_x}=\delta_{i,x(b)}
    \quad\mbox{with\ }\eta_x\in\rrh.
\end{equation}
The fact that $\eta_x$ lies in this real-linear subspace means that
the corresponding operator on $\hh$ is hermitian, or, expressed in
the standard basis that
\begin{equation}\label{etaHerm}
    \braket{\alpha\beta}{\eta_x}=\overline{\braket{\beta\alpha}{\eta_x}}.
\end{equation}
It is clear that if Alice can find any safe vectors at all, she has
some success at a {\it unambiguous retrodiction} game, in which she
is allowed to pass, but has to be absolutely sure of her guess
otherwise. As in the problem of ``unambiguous discrimination''
\cite{UAD} her aim would be to minimize the probability for pass
moves. In the mean king problem, however, her success probability is
required to be unity, which is the same as saying that
$\sum_xF_x=\idty$, and a guess $x\in X$ is produced in every run.

\subsection{Necessary conditions}
The Theorem states necessary conditions for the existence of a
strategy only in the tomographically complete case.  Then
$\rrh^\perp=\{0\}$, and the only choice Alice has is to pick safe
vectors, which are multiples of the $\eta_x$ as in (\ref{etax}).
This fixes the operators in (\ref{basiceq1}) to be
\begin{equation}\label{lammda}
    (\idty\otimes S)^*F_x(\idty\otimes S)=p(x)
    \kettbra{\eta_x},
\end{equation}
with $p(x)\geq0$. The values for $x$ not allowing a non-zero safe
vector can be subsumed by setting $p(x)=0$.  The overall
normalization condition $\sum_xF_x=\idty$ then reads
\begin{equation}\label{normalize}
    N=\sum_xp(x)\kettbra{\eta_x}=(\idty\otimes S^*S).
\end{equation}
Taking matrix elements of this equation in the standard basis and
using the hermiticity (\ref{etaHerm}), we find
\begin{equation}\label{NHerm}
    \bra{\alpha\beta}N\ket{\alpha'\beta'}
    =\overline{\bra{\beta\alpha}N\ket{\beta'\alpha'}}.
\end{equation}
By (\ref{normalize}) this amounts to $
\delta_{\alpha\alpha'}n_{\beta\beta'}
 =\delta_{\beta\beta'}\overline{n_{\alpha\alpha'}}$,
where $n$ is the matrix of $S^*S$. With $\alpha=\alpha'=1$ we find
that $S^*S$ is also a multiple of the identity matrix. From the
normalization condition $\tr\,S^*S=1$ this multiple must be $1/d$.
Since $S^*S$ is just the reduced density operator of the restricted
state, we have thus shown item (3) of the Theorem: In the
tomographically complete case, the initial state of Alice must be
maximally entangled.

The connection with classical models is seen by taking the matrix
elements of equation (\ref{normalize}) with other vectors. To begin
with, let us consider the matrix element with $\Omega$. Then, since
$N=(1/d)\idty$, and $\braket{\eta_x}{\Omega}=1$ whenever
$p(x)\neq0$, we get $\bra\Omega
N\ket\Omega=(1/d)\braket\Omega\Omega=1=\sum_xp(x)$. Hence the $p(x)$
must indeed be a probability distribution on $X$, which is the same
as the collection of all measurement outcomes. Furthermore, for any
bases $b,c$, and associated outcomes $i,j$,
\begin{equation}\label{Nmatel}
  \bra{\bsh bi}N\ket{\bsh cj} =
  \sum_xp(x)\delta_{i,x(b)}\delta_{j,x(c)}.
\end{equation}
Clearly, the right hand side is exactly the marginal $p_{bc}(i,j)$
of the probability distribution $p$ with respect to the $d$-valued
variables $b$ and $c$. On the other hand, since $N=(1/d)\idty$, the
left hand side evaluates to
 $(1/d)\braket{\bsh bi}{\bsh cj}=(1/d)|\braket{\bs bi}{\bs cj}|^2$,
so $p$ is exactly a classical model in the sense described in
Sect.~\ref{sec:results}.

This completes the proof of the Theorem, part (2), and the
corresponding statement in part (3).

\subsection{Sufficient conditions}
Let us now suppose, as in part (1) of the Theorem, that we are given
$k\leq(d+1)$ bases. Then, for each $x$, (\ref{etax}) is an
inhomogeneous linear system of equations for the vector $\eta_x$.
Taking only the first $d-1$ equations for each basis, plus one
normalization equation $\braket\Omega{\eta_x}=1$ eliminates the
trivial dependencies between these equations, so we have $k(d-1)+1$
equations for a vector in $\rrh$. The condition of non-degeneracy
described in Sect.~\ref{sec:results} is equivalent to saying that
all these equations are non-singular, hence under the hypothesis of
part (1) of the Theorem, $\eta_x$ exists for all $x$.

Now suppose that a classical model exists in the form of a set of
$p(x)\geq0$ such that the marginals (\ref{Nmatel}) are consistent
with $N=(1/d)\idty$. Note, however, that $\rrh$ may now be a proper
subspace, and the matrix elements with all $\bsh bi$ do not
determine the operator $N$ completely. Nevertheless, we can set
\begin{equation}\label{ftwiddle}
   F_x=d\,p(x)\, \kettbra{\eta_x} +\widetilde F_x
\end{equation}
with $\widetilde F_x\geq0$ summing to the projection onto
$\rrh^\perp$. Then it is immediate that with a maximally entangled
initial state, i.e., with the choice $S=(1/\sqrt d)\idty$, the basic
equation (\ref{basiceq1}) is satisfied. This completes the proof of
part (1) of the Theorem.

\

\section{Finding a strategy numerically}

Given the marginals of equation \eqref{probs}, the existence of a
classical model is a linear feasibility program in the $p(x)$. It
can also be cast as a semidefinite program, namely to maximize
$\sum_x p(x)$ subject to the constraints $p(x)\geq0$ and
 $\sum_x p(x)\ketbra{\eta_x}{\eta_x}  \leq \openone /d$.
If the maximum turns out to be $\sum_xp(x)=1$, we have found the
desired joint distribution. Otherwise, this is the probability for
Alice to find an answer in the unambiguous retrodiction game
described at the end of III.A. The following table lists the
numerical results for low dimensions with independent Haar
distributed bases, where $p_S$ is the probability that a safe
strategy exists, $E_S$ is the expected overall success probability
for unambiguous retrodiction, and $N$ is the sample size we used.

\begin{center}
\begin{tabular}{r|c|c|l}
  $d$& $p_S$ & $E_S$ & $\log_{10}N$ \\ \hline
  $2$& \quad.3334       & \quad .6666& \quad 7\\
  $3$& \quad.0013       & \quad .398 & \quad 6\\
  $4$& \quad 0          & \quad .34 & \quad 3\\
\end{tabular}
\end{center}

Higher dimensions, with $d^{d+1}$ variables and constraints, are a
serious challenge for PC based computation. For $d=6$, a strategy
rarely exists, but one can first ``debias'' the bases with a
gradient search minimizing $\sum_{i,j,a,b}p_{ab}(i,j)^2$. Instead of
a semidefinite program one can then use the so-called EM-algorithm
\cite{EM,Gill} to find a joint distribution, and this is typically
successful for the debiased case, although convergence is rather
slow.

\noindent{\it Acknowledgements:}\/ We thank Richard~Gill for
valuable discussions. This work has been supported by the DFG and JOMC, Braunschweig.


\end{document}